\documentclass[10pt]{article}
\usepackage{graphicx}
\usepackage{amsmath}
\usepackage{amssymb}
\usepackage{caption2}
\begin{document}
\begin{center}
{\large {\bf \sc{  Analysis of the decay $Y(4500)\to D^*\bar{D}^*\pi$  with the light-cone QCD sum rules  }}} \\[2mm]
Zhi-Gang  Wang \footnote{E-mail: zgwang@aliyun.com.  }     \\
 Department of Physics, North China Electric Power University, Baoding 071003, P. R. China
\end{center}

\begin{abstract}
In this work, we tentatively assign the $Y(4500)$ as the $[uc]_{\tilde{A}}[\overline{uc}]_{V}+[uc]_{V}[\overline{uc}]_{\tilde{A}}+[dc]_{\tilde{A}}[\overline{dc}]_{V}
+[dc]_{V}[\overline{dc}]_{\tilde{A}}$ tetraquark state with the quantum numbers $J^{PC}=1^{--}$, and   study the three-body strong decay $Y(4500)\to D^{*-}D^{*0}\pi^+$ with the light-cone QCD sum rules. It is the first time to use the light-cone QCD sum rules to calculate the four-hadron coupling constants, the approach can be extended to study other three-body strong decays directly  and diagnose  the $X$, $Y$ and $Z$ states.
 \end{abstract}

 PACS number: 12.39.Mk, 12.38.Lg

Key words: Tetraquark  state, QCD sum rules

\section{Introduction}
In last two decays,  several vector  charmonium-like states have been observed,  they  cannot be accommodated comfortably  in the traditional  charmonia,  we have to introduce additional quark or gluon degrees of freedom in assignments  \cite{PDG}.
For example, the $Y(4260)$ observed in the $J/\psi \pi^+\pi^-$ invariant  mass spectrum by the BaBar collaboration \cite{BaBar4260-0506}, the $Y(4220)$ and $Y(4390)$ ($Y(4320)$) observed in the $h_c \pi^+\pi^-$ ($J/\psi \pi^+\pi^-$) invariant mass spectrum by the BESIII collaboration  \cite{BES-Y4390,BES-Y4220-Y4320}, and the $Y(4360)$ and $Y(4660)$ ($Y(4630)$) observed in the $\psi^\prime \pi^+\pi^-$ ($\Lambda_c^+\bar{\Lambda}_c^-$) invariant mass spectrum by the
  Belle collaboration \cite{Belle4660-0707-1,Belle4660-0707-2,Belle4630-0807} are excellent candidates for the vector tetraquark states.

In 2022, the BESIII collaboration explored the $e^+e^-\to K^+K^-J/\psi$ cross sections    at center-of-mass energies from 4.127 to 4.600 GeV  based on $15.6\,\rm{fb}^{-1}$ data, and observed two resonant structures, one is consistent with the established $Y(4230)$; the other was  observed for the first time with a significance larger than $8\sigma$ and denoted as $Y(4500)$, its Breit-Wigner  mass and width are $4484.7\pm 13.3\pm 24.1\,\rm{MeV}$ and $111.1\pm 30.1\pm 15.2\, \rm{MeV}$, respectively \cite{BESIII-KK-4500}.

Recently, the BESIII collaboration explored the Born cross sections of the process  $e^+e^-\to D^{*-}D^{*0}\pi^+$ at center-of-mass energies from 4.189 to 4.951 GeV using the data samples corresponding  to an integrated luminosity of $17.9\,\rm{fb}^{-1}$ and observed three enhancements, whose  masses are $4209.6\pm 4.7\pm 5.9\,\rm{MeV}$, $4469.1\pm26.2\pm3.6\,\rm{MeV}$ and $4675.3\pm 29.5\pm 3.5\,\rm{MeV}$, respectively,  and widths are $81.6\pm 17.8\pm 9.0\,\rm{MeV}$, $246.3\pm 36.7\pm 9.4\,\rm{MeV}$ and $218.3\pm 72.9\pm 9.3\,\rm{MeV}$, respectively. The first and third resonances are consistent with the $Y(4230)$ and $Y(4660)$ states, respectively, while the second resonance is compatible with the $Y(4500)$ \cite{X4500-BESIII}.

In fact, analogous decays were already observed  in the process $e^+e^-\to Y \to \pi^+ D^0D^{*-}$ for the center-of-mass energies from 4.05 to 4.60 GeV by the BESIII collaboration in 2018, and  the two enhancements $Y$ lie  around 4.23 and 4.40 GeV, respectively \cite{BESIII-DDvpi}.

In the scenario of tetraquark states, the calculations based on the QCD sum rules have given several    reasonable assignments of the $Y$ states \cite{Nielsen-4260-4460,ChenZhu-Vector-Axial,WangY4360Y4660-1803,Wang-tetra-formula,WangEPJC-1601-Mc,ZhangHuang-PRD,
ZhangHuang-JHEP,Vector-Tetra-WZG-P-wave-1,Vector-Tetra-WZG-P-wave,Vector-Tetra-WZG-4100,Vector-4660-Azizi,
WZG-Vector-NPB}. For example, in Ref.\cite{WZG-Vector-NPB}, we take the scalar, pseudoscalar, axialvector, vector and tensor  (anti)diquarks   to construct  vector and tensor  four-quark currents without introducing explicit P-waves,  and explore the mass spectrum of the vector hidden-charm tetraquark states via the QCD sum rules in a comprehensive way. At the energy about $4.5\,\rm{GeV}$, we obtain three hidden-charm tetraquark states with the $J^{PC}=1^{--}$, the tetraquark states with the symbolic structures  $[uc]_{\tilde{V}}[\overline{dc}]_{A}-[uc]_{A}[\overline{dc}]_{\tilde{V}}$,
 $[uc]_{\tilde{A}}[\overline{dc}]_{V}+[uc]_{V}[\overline{dc}]_{\tilde{A}}$ and
 $[uc]_{S}[\overline{dc}]_{\tilde{V}}-[uc]_{\tilde{V}}[\overline{dc}]_{S}$ have
 the masses $4.53\pm0.07\, \rm{GeV}$, $4.48\pm0.08\,\rm{GeV}$ and $4.50\pm0.09\,\rm{GeV}$, respectively \cite{WZG-Vector-NPB}.
  Thus we have three candidates for the $Y(4500)$, the best assignment of the symbolic structure  is $[uc]_{\tilde{A}}[\overline{dc}]_{V}+[uc]_{V}[\overline{dc}]_{\tilde{A}}=Y(4500)$ comparing  with the BESIII experimental data $M_{Y(4500)}=4469.1\pm26.2\pm3.6\,\rm{MeV}$ \cite{X4500-BESIII}, where we have taken the isospin limit, the tetraquark states with the  valence quark   structures,
 \begin{eqnarray}
 \bar{c}c\bar{d}u, \, \, \bar{c}c\bar{u}d, \, \, \bar{c}c\frac{\bar{u}u-\bar{d}d}{\sqrt{2}}, \, \, \bar{c}c\frac{\bar{u}u+\bar{d}d}{\sqrt{2}}\, ,
 \end{eqnarray}
 have degenerated  masses and pole residues.
   As there exist three four-quark currents with the $J^{PC}=1^{--}$, which couple potentially
to the vector hidden-charm tetraquark states with almost degenerated masses \cite{WZG-Vector-NPB}, we can also tentatively  say that there only exists  one vector tetraquark state with three different Fock components $[uc]_{\tilde{V}}[\overline{dc}]_{A}-[uc]_{A}[\overline{dc}]_{\tilde{V}}$,
 $[uc]_{\tilde{A}}[\overline{dc}]_{V}+[uc]_{V}[\overline{dc}]_{\tilde{A}}$ and
 $[uc]_{S}[\overline{dc}]_{\tilde{V}}-[uc]_{\tilde{V}}[\overline{dc}]_{S}$, we can choose either Fock component (in other words, current) to explore the hadronic properties. At the first step, we choose the optimal current corresponding to the optimal Fock component $[uc]_{\tilde{A}}[\overline{dc}]_{V}+[uc]_{V}[\overline{dc}]_{\tilde{A}}$, i.e. the current $J_{\mu}^{Y}(0)$ in Eq.\eqref{current-JY} in the isospin limit.
However, we cannot assign a hadron unambiguously with the mass alone, we have to explore  the decay width to make more robust assignment. If we want to investigate the three-body strong decays  $Y\to J/\psi \pi^+\pi^-$, $\psi^\prime \pi^+\pi^-$, $J/\psi K^+K^-$, $h_c \pi^+\pi^-$, $D^0D^{*-}\pi^+$ and $D^{*-}D^{*0}\pi^+$ with the QCD sum rules directly, we have to introduce four-point correlation functions, the hadronic spectral densities are complex enough to destroy the reliability of the calculations.

In this work, we tentatively assign the $Y(4500)$ as the $[uc]_{\tilde{A}}[\overline{uc}]_{V}+[uc]_{V}[\overline{uc}]_{\tilde{A}}+[dc]_{\tilde{A}}[\overline{dc}]_{V}
+[dc]_{V}[\overline{dc}]_{\tilde{A}}$ tetraquark state with the $J^{PC}=1^{--}$, and   extend our previous works to study the three-body strong decay $Y(4500)\to D^{*-}D^{*0}\pi^+$ with the light-cone QCD sum rules, where only three-point correlation function is needed. It is the first time to investigate the three-body strong decays with the light-cone QCD sum rules.  In our previous works, we have obtained rigorous quark-hadron duality for the
three-point correlation functions, which work very well.
There are other procedures in dealing with the three-point QCD sum rules exploring  the hadronic coupling constants \cite{Nilesen-decay,Chen-Chen-decay,Azizi-decay}, for detailed discussions about the differences, one can consult Refs.\cite{WZG-ZJX-Zc-Decay,WZG-Y4660-Decay}.

The article is arranged as follows:  we derive the light-cone QCD sum rules for the  $YD^*\bar{D}^*\pi$ coupling constants in section 2; in section 3, we present numerical results and discussions; section 4 is reserved for our conclusion.

\section{Light-cone QCD sum rules for  the  $YD^*\bar{D}^*\pi$ coupling constants}
\begin{figure}
\centering
   \includegraphics[totalheight=4cm,width=6cm]{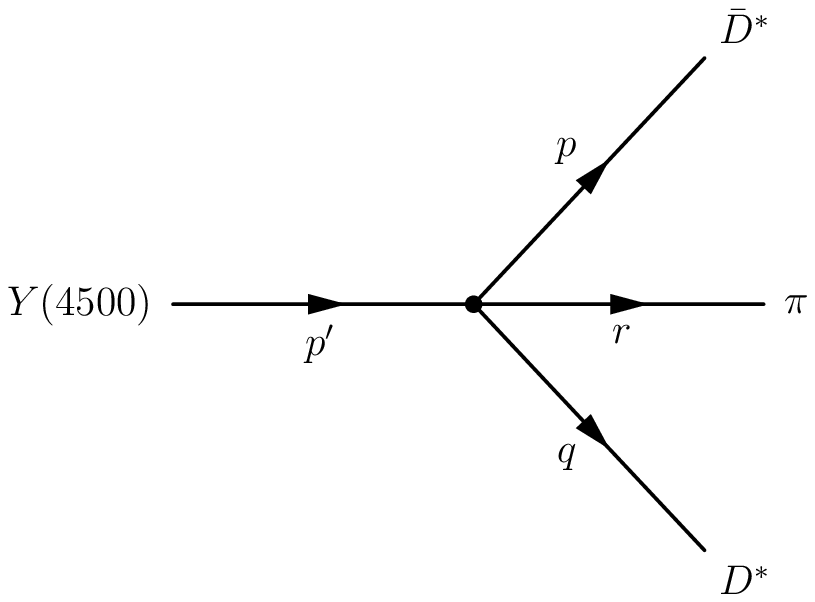}
   \caption{ The decay $Y(4500)\to \bar{D}^{*}D^{*}\pi^+$.}\label{Y4500-vertex-fig}
\end{figure}

Firstly, we write down  the three-point correlation function  $\Pi_{\mu\alpha\beta}(p)$ in the light-cone QCD sum rules,
\begin{eqnarray}
\Pi_{\mu\alpha\beta}(p,q)&=&i^2\int d^4xd^4y \, e^{-ip\cdot x}e^{-iq\cdot y}\, \langle 0|T\left\{J_{\mu}^{Y}(0)J_\alpha^{D^{*+}}(x)J^{\bar{D}^{*0}}_{\beta}(y)\right\}|\pi(r)\rangle\, ,
\end{eqnarray}
where the currents
\begin{eqnarray}\label{current-JY}
J_{\mu}^{Y}(0)&=&\frac{\varepsilon^{ijk}\varepsilon^{imn}}{2}\Big[u^{T}_j(0)C\sigma_{\mu\nu}\gamma_5 c_k(0)\bar{u}_m(0)\gamma_5\gamma^\nu C \bar{c}^{T}_n(0)+u^{T}_j(0)C\gamma^\nu\gamma_5 c_k(0)\bar{u}_m(0)\gamma_5\sigma_{\mu\nu} C \bar{c}^{T}_n(0) \nonumber\\
&&+d^{T}_j(x)C\sigma_{\mu\nu}\gamma_5 c_k(0)\bar{d}_m(0)\gamma_5\gamma^\nu C \bar{c}^{T}_n(0)+d^{T}_j(0)C\gamma^\nu\gamma_5 c_k(0)\bar{d}_m(0)\gamma_5\sigma_{\mu\nu} C \bar{c}^{T}_n(0) \Big] \, , \nonumber\\
J_{\alpha}^{D^{*+}}(y)&=&\bar{d}(x)\gamma_{\alpha} c(x) \, ,\nonumber \\
J_{\beta}^{\bar{D}^{*0}}(x)&=&\bar{c}(y)\gamma_{\beta} u(y) \, ,
\end{eqnarray}
interpolate the mesons $Y(4500)$,  $\bar{D}^*$ and $D^*$, respectively \cite{WZG-Vector-NPB}, the $|\pi(r)\rangle$ is the external $\pi$ state. The physical process is shown explicitly in Fig.\ref{Y4500-vertex-fig}.

In the present work, we take the isospin limit, the current $J_{\mu}^{Y}(x)$ in Eq.\eqref{current-JY} and the current $J_{-,\mu}^{\widetilde{A}V}(x)$
chosen in Ref.\cite{WZG-Vector-NPB} couple potentially to the vector tetraquark states with the same masses and pole residues,
where
\begin{eqnarray}
J_{-,\mu}^{\widetilde{A}V}(x)&=&\frac{\varepsilon^{ijk}\varepsilon^{imn}}{\sqrt{2}}\Big[u^{Tj}(x)C\sigma_{\mu\nu}\gamma_5 c^k(x)\bar{d}^m(x)\gamma_5\gamma^\nu C \bar{c}^{Tn}(x)\nonumber\\
&&+u^{Tj}(x)C\gamma^\nu\gamma_5 c^k(x)\bar{d}^m(x)\gamma_5\sigma_{\mu\nu} C \bar{c}^{Tn}(x) \Big]\, .
\end{eqnarray}

At the hadron side, we insert  a complete set of intermediate hadronic states having non-vanishing couplings  with the interpolating currents into the three-point correlation function, and  isolate the ground state contributions clearly,
\begin{eqnarray}\label{Hadron-CT}
\Pi_{\mu\alpha\beta}(p,q)&=& \lambda_Y f_{D^*}^2M_{D^*}^2 \frac{-iG_{\pi}r_\tau+iG_{Y}p^\prime_\tau}{(M_{Y}^2-p^{\prime2})(M_{\bar{D}^*}^2-p^2)(M_{D^*}^2-q^2)}\varepsilon^{\rho\sigma\lambda\tau}  \left( -g_{\mu\rho}+\frac{p^\prime_\mu p^\prime_\rho}{p^{\prime2}}\right)\nonumber\\
&&\left( -g_{\alpha\sigma}+\frac{p_\alpha p_\sigma}{p^2}\right)\left( -g_{\lambda\beta}+\frac{q_\lambda q_\beta}{q^2}\right) + \cdots\, ,
\end{eqnarray}
where $p^\prime=p+q+r$, the  decay constants $\lambda_Y$, $f_{D^*}$, $f_{\bar{D}^*}$ and hadronic coupling constants $G_\pi$, $G_Y$ are defined by,
\begin{eqnarray}
\langle0|J_{\mu}^{Y}(0)|Y_c(p^\prime)\rangle&=&\lambda_{Y} \varepsilon_{\mu} \,\, , \nonumber \\
\langle0|J_{\alpha}^{D^*\dag}(0)|\bar{D}^*(p)\rangle&=&f_{\bar{D}^*} M_{\bar{D}^*} \xi_\alpha \,\, , \nonumber \\
\langle0|J_{\beta}^{\bar{D}^{*\dag}}(0)|D^*(q)\rangle&=&f_{D^*} M_{D^*} \zeta_\beta \,\, ,
\end{eqnarray}
\begin{eqnarray}\label{define-G-pi-X}
\langle Y_c(p^\prime)|\bar{D}^*(p)D^*(q)\pi(r)\rangle&=&G_{\pi}\varepsilon^{\rho\sigma\lambda\tau}\varepsilon^*_\rho\xi_\sigma\zeta_\lambda r_\tau-G_{Y}\varepsilon^{\rho\sigma\lambda\tau}\varepsilon^*_\rho\xi_\sigma\zeta_\lambda p^\prime_\tau \,\, ,
\end{eqnarray}
 the $\varepsilon_{\mu}$, $\xi_\alpha$ and $\zeta_\beta$  are polarization vectors of the $Y(4500)$, $\bar{D}^*$ and  $D^*$, respectively. In the isospin limit, $m_u=m_d$, $f_{D^*}=f_{\bar{D}^*}$ and $M_{D^*}=M_{\bar{D}^*}$.

We multiply Eq.\eqref{Hadron-CT} with the tensor $\varepsilon_{\theta\omega}{}^{\alpha\beta}$ and obtain
\begin{eqnarray}\label{Hadron-CT-W}
\widetilde{\Pi}_{\mu\theta\omega}(p,q)&=& \varepsilon_{\theta\omega}{}^{\alpha\beta}\,\Pi_{\mu\alpha\beta}(p,q)\nonumber\\
&=&\lambda_Y f_{D^*}^2M_{D^*}^2 \frac{iG_{\pi}\left(g_{\mu\omega}r_\theta-g_{\mu\theta}r_\omega\right)-iG_{Y}
\left(g_{\mu\omega}p^\prime_\theta-g_{\mu\theta}p^\prime_\omega\right)}{(M_{Y}^2-p^{\prime2})(M_{\bar{D}^*}^2-p^2)
(M_{D^*}^2-q^2)}  + \cdots\, .
\end{eqnarray}
Again, we take the isospin limit,  then $\widetilde{\Pi}_{\mu\theta\omega}(p,q)=\widetilde{\Pi}_{\mu\theta\omega}(q,p)$, such a relation can simplify the calculations at the QCD side greatly, and we write down the relevant components,
\begin{eqnarray}
\widetilde{\Pi}_{\mu\theta\omega}(p,q)&=&\left[i\Pi_{\pi}(p^{\prime2},p^2,q^2)-i\Pi_{Y}(p^{\prime2},p^2,q^2)\right]
\left(g_{\mu\omega}r_\theta-g_{\mu\theta}r_\omega\right) \nonumber\\
&&+i\Pi_{Y}(p^{\prime2},p^2,q^2)
\left(g_{\mu\omega}q_\theta-g_{\mu\theta}q_\omega\right)+\cdots \, ,
\end{eqnarray}
where
\begin{eqnarray}
\Pi_{\pi}(p^{\prime2},p^2,q^2)&=& \frac{\lambda_Y f_{D^*}^2M_{D^*}^2 G_{\pi}}{(M_{Y}^2-p^{\prime2})(M_{\bar{D}^*}^2-p^2)
(M_{D^*}^2-q^2)}  + \cdots\, ,\nonumber\\
\Pi_{Y}(p^{\prime2},p^2,q^2)&=& \frac{\lambda_Y f_{D^*}^2M_{D^*}^2 G_{Y}}{(M_{Y}^2-p^{\prime2})(M_{\bar{D}^*}^2-p^2)
(M_{D^*}^2-q^2)}  + \cdots\, .
\end{eqnarray}
Then we choose the tensor structures $g_{\mu\omega}r_\theta-g_{\mu\theta}r_\omega$ and $g_{\mu\omega}q_\theta-g_{\mu\theta}q_\omega$ to study the hadronic coupling constants  $G_\pi$ and $G_{Y}$, respectively. And  we obtain the hadronic  spectral densities $\rho_H(s^\prime,s,u)$ through triple  dispersion relation,
\begin{eqnarray}
\Pi_{H}(p^{\prime2},p^2,q^2)&=&\int_{\Delta_s^{\prime2}}^\infty ds^{\prime} \int_{\Delta_s^2}^\infty ds \int_{\Delta_u^2}^\infty du \frac{\rho_{H}(s^\prime,s,u)}{(s^\prime-p^{\prime2})(s-p^2)(u-q^2)}\, ,
\end{eqnarray}
where the $\Delta_{s}^{\prime2}$, $\Delta_{s}^{2}$ and
$\Delta_{u}^{2}$ are the thresholds, and we add the subscript $H$ to represent the hadron side.

\begin{figure}
 \centering
  \includegraphics[totalheight=5cm,width=12cm]{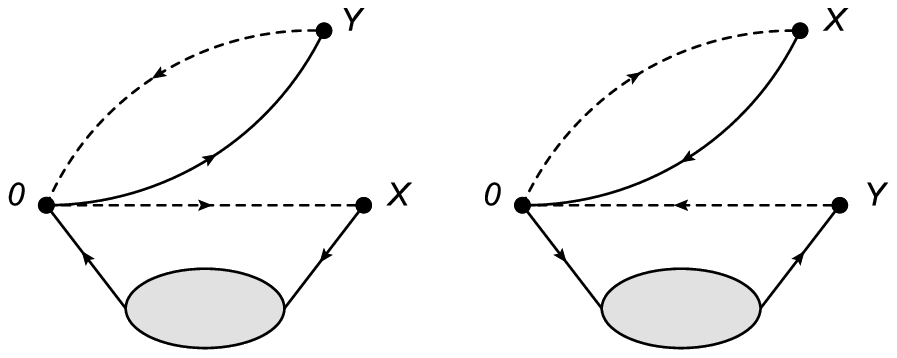}
 \caption{ The lowest order  Feynman diagrams, where the dashed (solid) lines denote the heavy (light) quark lines, the ovals denote the external $\pi^+$ meson.}\label{Y-DvDvpi-fig}
\end{figure}

We carry out   the operator product expansion up to the vacuum condensates of dimension 5 and neglect the tiny gluon condensate contributions \cite{WZG-ZJX-Zc-Decay,WZG-Y4660-Decay},
\begin{eqnarray}\label{QCD-CT-1}
\Pi_{\pi}(p^2,q^{\prime2},q^2)&=& f_\pi m_c \int_0^1 du \varphi_\pi(u) \left[\int_0^1 dx x\bar{x}\frac{\Gamma(\epsilon-1)}{2\pi^2(p^2-\tilde{m}_c^2)^{\epsilon-1}}
-\frac{2m_c\langle\bar{q}q\rangle}{3(p^2-m_c^2)}\right.\nonumber\\
&&\left.+\frac{m_c^3\langle\bar{q}g_s\sigma Gq\rangle}{3(p^2-m_c^2)^3} \right]\frac{1}{(q+ur)^2-m_c^2}\nonumber\\
&&+ \frac{f_\pi m_\pi^2}{m_u+m_d} \int_0^1 du \varphi_5(u)\bar{u} \left[\int_0^1 dx x\bar{x}\frac{\Gamma(\epsilon-1)}{2\pi^2(p^2-\tilde{m}_c^2)^{\epsilon-1}}
-\frac{2m_c\langle\bar{q}q\rangle}{3(p^2-m_c^2)}\right. \nonumber\\
&&\left.+\frac{m_c^3\langle\bar{q}g_s\sigma Gq\rangle}{3(p^2-m_c^2)^3} \right]\frac{1}{(q+ur)^2-m_c^2}\nonumber\\
&&-\frac{f_\pi m_c^2\langle\bar{q}g_s\sigma Gq\rangle}{36} \int_0^1 du \varphi_\pi(u) \frac{1}{(p^2-m_c^2)((q+ur)^2-m_c^2)^2} \nonumber\\
&&+\frac{f_\pi m_\pi^2 m_c\langle\bar{q}g_s\sigma Gq\rangle}{36(m_u+m_d)} \int_0^1 du \varphi_5(u)\bar{u} \frac{1}{(p^2-m_c^2)((q+ur)^2-m_c^2)^2} \, ,
\end{eqnarray}

\begin{eqnarray}\label{QCD-CT-2}
\Pi_{Y}(p^2,q^{\prime2},q^2)&=& \frac{f_\pi m_\pi^2}{m_u+m_d} \int_0^1 du \varphi_5(u) \left[\int_0^1 dx x\bar{x}\frac{\Gamma(\epsilon-1)}{2\pi^2(p^2-\tilde{m}_c^2)^{\epsilon-1}}
-\frac{2m_c\langle\bar{q}q\rangle}{3(p^2-m_c^2)}\right. \nonumber\\
&&\left.+\frac{m_c^3\langle\bar{q}g_s\sigma Gq\rangle}{3(p^2-m_c^2)^3} \right]\frac{1}{(q+ur)^2-m_c^2}\nonumber\\
&&+\frac{f_\pi m_\pi^2 m_c\langle\bar{q}g_s\sigma Gq\rangle}{36(m_u+m_d)} \int_0^1 du \varphi_5(u) \frac{1}{(p^2-m_c^2)((q+ur)^2-m_c^2)^2} \nonumber\\
&&+f_{3\pi}m_\pi^2 \int_0^1 dx \bar{x} \left[\frac{3\Gamma(\epsilon)}{8\pi^2(p^2-\tilde{m}_c^2)^\epsilon}-\frac{p^2}{2\pi^2(p^2-\tilde{m}_c^2)}  \right]\frac{1}{q^2-m_c^2} \nonumber\\
&&-f_{3\pi}m_\pi^2 \int_0^1 dx x\bar{x} \left[\frac{\Gamma(\epsilon-1)}{2\pi^2(p^2-\tilde{m}_c^2)^{\epsilon-1}}+\frac{p^2\Gamma(\epsilon)}{2\pi^2(p^2-\tilde{m}_c^2)^{\epsilon}}  \right]\frac{1}{(q^2-m_c^2)^2} \nonumber\\
&&-f_{3\pi}m_\pi^2 \int_0^1 dx x \left[\frac{3\Gamma(\epsilon)}{8\pi^2(p^2-\tilde{m}_c^2)^\epsilon}+\frac{p^2}{4\pi^2(p^2-\tilde{m}_c^2)}  \right]\frac{1}{q^2-m_c^2} \, ,
\end{eqnarray}
where $q^\prime=q+r$, $\bar{u}=1-u$, $\bar{x}=1-x$, $\tilde{m}_c^2=\frac{m_c^2}{x}$, $(q-ur)^2-m_c^2=(1-u)q^2+u(q+r)^2-u\bar{u}m_\pi^2-m_c^2$.
And we have used the definitions for the $\pi$ light-cone distribution functions \cite{PBall-LCDF},
\begin{eqnarray}
\langle 0|\bar{d}(0)\gamma_\mu\gamma_5 u(x)|\pi(r)\rangle &=&if_\pi r_\mu \int_0^1 du e^{-iur\cdot x} \varphi_{\pi}(u)+\cdots\, , \nonumber\\
\langle 0|\bar{d}(0)\sigma_{\mu\nu}\gamma_5 u(x)|\pi(r)\rangle &=&\frac{i}{6}\frac{f_\pi m_\pi^2}{m_u+m_d} \left(r_\mu x_\nu -r_\nu x_\mu \right) \int_0^1 du e^{-iur\cdot x} \varphi_{\sigma}(u) \, , \nonumber\\
\langle 0|\bar{d}(0)i\gamma_5 u(x)|\pi(r)\rangle &=& \frac{f_\pi m_\pi^2}{m_u+m_d}  \int_0^1 du e^{-iur\cdot x} \varphi_{5}(u) \, ,
\end{eqnarray}
and the approximation,
\begin{eqnarray} \label{qGq}
\langle 0|\bar{d}(x_1)\sigma_{\mu\nu}\gamma_5g_sG_{\alpha\beta}(x_2) u(x_3)|\pi(r)\rangle &=&if_{3\pi}\left( r_\mu r_\alpha g_{\nu\beta}+r_\nu r_\beta g_{\mu\alpha}-r_\nu r_\alpha g_{\mu\beta}-r_\mu r_\beta g_{\nu\alpha}\right) \, ,
\end{eqnarray}
for the twist-3 quark-gluon light-cone distribution functions.
Such terms proportional to $m_\pi^2$ and their contributions are greatly suppressed \cite{WZG-SLWan,WZG-Ds2460}, the approximation in Eq.\eqref{qGq} works well. However, the terms proportional to $\frac{m_\pi^2}{m_u+m_d}$ are Chiral enhanced due to the Gell-Mann-Oakes-Renner relation $\frac{f_\pi m_\pi^2}{m_u+m_d}=-\frac{2\langle\bar{q}q\rangle}{f_\pi}$, and we take account of those contributions fully in Eqs.\eqref{QCD-CT-1}-\eqref{QCD-CT-2}. In the following, we list out the light-cone distribution functions explicitly,
\begin{eqnarray}
\varphi_\pi(u)&=&6u\bar{u}\left[1+A_2\frac{3}{2}\left(5t^2-1 \right)+A_4 \frac{15}{8}\left(21t^4-14t^2+1 \right)  \right]\, ,\nonumber\\
\varphi_5(u)&=&1+B_2\frac{1}{2}\left(3t^2-1 \right)+B_4 \frac{1}{8}\left(35t^4-30t^2+3 \right)  \, ,\nonumber\\
\varphi_\sigma(u)&=&6u\bar{u}\left[1+C_2\frac{3}{2}\left(5t^2-1 \right) \right]\, ,
\end{eqnarray}
where $t=2u-1$, and the coefficients $A_2=0.44$, $A_4=0.25$, $B_2=0.43$, $B_4=0.10$, $C_2=0.09$, and the decay constant $f_{3\pi}=0.0035\,\rm{GeV}^2$ at the energy scale $\mu=1\,\rm{GeV}$ \cite{PBall-LCDF,Braun-f3pi}.
In the present work, we neglect the twist-4 light-cone distribution functions due to their small contributions. In Fig.\ref{Y-DvDvpi-fig}, we draw the lowest order Feynman diagrams as example to illustrate the operator product expansion.

In the soft limit $r_\mu \to 0$, $(q+r)^2=q^2$, we can set $\Pi_{\pi/Y}(p^2,q^{\prime2},q^2)=\Pi_{\pi/Y}(p^2,q^2)$, then we obtain the QCD spectral densities $\rho_{QCD}(s,u)$  through double dispersion relation,
\begin{eqnarray}
\Pi^{QCD}_{\pi/Y}(p^2,q^2)&=& \int_{\Delta_s^2}^\infty ds \int_{\Delta_u^2}^\infty du \frac{\rho_{QCD}(s,u)}{(s-p^2)(u-q^2)}\, ,
\end{eqnarray}
again the $\Delta_s^2$ and $\Delta_u^2$  are the thresholds, we add the superscript or subscript $QCD$ to stand for  the QCD side.

We match the hadron side with the QCD side  bellow the continuum thresholds $s_0$ and $u_0$ to acquire   rigorous quark-hadron  duality  \cite{WZG-ZJX-Zc-Decay,WZG-Y4660-Decay},
 \begin{eqnarray}
  \int_{\Delta_s^2}^{s_{0}}ds \int_{\Delta_u^2}^{u_0}du  \frac{\rho_{QCD}(s,u)}{(s-p^2)(u-q^2)}&=& \int_{\Delta_s^2}^{s_0}ds \int_{\Delta_u^2}^{u_0}du  \left[ \int_{\Delta_{s}^{\prime2}}^{\infty}ds^\prime  \frac{\rho_H(s^\prime,s,u)}{(s^\prime-p^{\prime2})(s-p^2)(u-q^2)} \right]\, ,
\end{eqnarray}
and  we carry out  the integral over $ds^\prime$ firstly,
then
\begin{eqnarray}
\Pi_{H}(p^{\prime2},p^2,q^2)&=& \frac{\lambda_Y f_{D^*}^2M_{D^*}^2 G_{\pi/Y}}{(M_{Y}^2-p^{\prime2})(M_{\bar{D}^*}^2-p^2)
(M_{D^*}^2-q^2)}  +\int_{s^\prime_0}^{\infty}ds^\prime\frac{\tilde{\rho}_{H}(s^\prime,M_{\bar{D}^*}^2,M_{D^*}^2)}
{(s^\prime-p^{\prime2})(M_{\bar{D}^*}^2-p^2)
(M_{D^*}^2-q^2)}\nonumber\\
&& +\cdots\, ,\nonumber\\
&=& \frac{\lambda_Y f_{D^*}^2M_{D^*}^2 G_{\pi/Y}}{(M_{Y}^2-p^{\prime2})(M_{\bar{D}^*}^2-p^2)
(M_{D^*}^2-q^2)}  +\frac{C_{\pi/Y}}{(M_{\bar{D}^*}^2-p^2)(M_{D^*}^2-q^2)}+\cdots\, ,
\end{eqnarray}
where $\rho_{H}(s^\prime,s,u)=\tilde{\rho}_{H}(s^\prime,s,u)\delta(s-M_{\bar{D}^*}^2)\delta(u-M_{D^*}^2)$,
and we introduce the parameters $C_{\pi/Y}$ to parameterize the contributions concerning  the higher resonances and continuum states in the $s^\prime$ channel,
\begin{eqnarray}
C_{\pi/Y}&=&\int_{s^\prime_0}^{\infty}ds^\prime\frac{\tilde{\rho}_{H}(s^\prime,M_{\bar{D}^*}^2,M_{D^*}^2)}{
s^\prime-p^{\prime2}}\, .
\end{eqnarray}
As the strong interactions among the ground states $\pi$, $D^*$, $\bar{D}^*$ and excited $Y^\prime$ states are complex, and we have no knowledge about the corresponding four-hadron contact vertex. In practical calculations, we can take the unknown functions $C_{\pi/Y}$ as free parameters and adjust the values to acquire flat platforms  for the hadronic coupling constants $G_{\pi/Y}$ with variations of the Borel parameters. Such a method works well in the case of three-hadron  contact vertexes \cite{WZG-ZJX-Zc-Decay,WZG-Y4660-Decay,WZG-X4140-decay,
WZG-X4274-decay,WZG-Z4600-decay,WZG-Zcs3985-decay,WZG-Zcs4123-decay}, and we expect it also works in the present work.

In Eq.\eqref{Hadron-CT} and Eq.\eqref{Hadron-CT-W}, there exist three poles in the limit
$p^{\prime2} \to M_{Y}^2$, $p^2 \to M_{\bar{D}^*}^2$ and $q^2 \to M_{D^*}^2$. According to the relation $M_Y\approx M_{\bar{D}^*}+M_{D^*}$, we can set $p^{\prime2}=4q^2$ in the correlation functions $\Pi_H(p^{\prime 2},p^2,q^2)$, and  perform  double Borel transform in regard  to the variables $P^2=-p^2$ and $Q^2=-q^2$ respectively, then we set the Borel parameters  $T_1^2=T_2^2=T^2$  to acquire   two QCD sum rules,
\begin{eqnarray} \label{pi-SR}
&&\frac{\lambda_{YD^*D^*}G_{\pi}}{4\left(\widetilde{M}_{Y}^2-M_{D^*}^2\right)} \left[ \exp\left(-\frac{M_{D^*}^2}{T^2} \right)-\exp\left(-\frac{\widetilde{M}_{Y}^2}{T^2} \right)\right]\exp\left(-\frac{M_{\bar{D}^*}^2}{T^2} \right)+C_{\pi} \exp\left(-\frac{M_{D^*}^2+M_{\bar{D}^*}^2}{T^2}  \right) \nonumber\\
&&= f_\pi m_c \int_{m_c^2}^{s_0}ds\int_0^1 du \varphi_\pi(u) \left[\frac{1}{2\pi^2}\int_{x_i}^1 dx x\bar{x}(s-\tilde{m}_c^2)
-\left(\frac{2m_c\langle\bar{q}q\rangle}{3}-\frac{m_c^3\langle\bar{q}g_s \sigma Gq\rangle}{6T^4}\right)\delta(s-m_c^2)\right]\nonumber\\
&&\exp\left(-\frac{s+m_c^2+u\bar{u}m_\pi^2}{T^2} \right)\nonumber\\
&&+ \frac{f_\pi m_\pi^2}{m_u+m_d}\int_{m_c^2}^{s_0}ds \int_0^1 du \varphi_5(u)\bar{u} \left[\frac{1}{2\pi^2}\int_{x_i}^1 dx x\bar{x}(s-\tilde{m}_c^2)
-\left(\frac{2m_c\langle\bar{q}q\rangle}{3}-\frac{m_c^3\langle\bar{q}g_s \sigma Gq\rangle}{6T^4}\right)\delta(s-m_c^2)\right]\nonumber\\
&&\exp\left(-\frac{s+m_c^2+u\bar{u}m_\pi^2}{T^2} \right)+\frac{f_\pi m_c^2\langle\bar{q}g_s\sigma Gq\rangle}{36T^2} \int_0^1 du \varphi_\pi(u) \exp\left(-\frac{2m_c^2+u\bar{u}m_\pi^2}{T^2} \right) \nonumber\\
&&-\frac{f_\pi m_\pi^2 m_c\langle\bar{q}g_s\sigma Gq\rangle}{36(m_u+m_d)T^2} \int_0^1 du \varphi_5(u)\bar{u} \exp\left(-\frac{2m_c^2+u\bar{u}m_\pi^2}{T^2} \right) \, ,
\end{eqnarray}

\begin{eqnarray} \label{X-SR}
&&\frac{\lambda_{YD^*D^*}G_{Y}}{4\left(\widetilde{M}_{Y}^2-M_{D^*}^2\right)} \left[ \exp\left(-\frac{M_{D^*}^2}{T^2} \right)-\exp\left(-\frac{\widetilde{M}_{Y}^2}{T^2} \right)\right]\exp\left(-\frac{M_{\bar{D}^*}^2}{T^2} \right)+C_{Y} \exp\left(-\frac{M_{D^*}^2+M_{\bar{D}^*}^2}{T^2}  \right) \nonumber\\
&&= \frac{f_\pi m_\pi^2}{m_u+m_d} \int_{m_c^2}^{s_0}ds\int_0^1 du \varphi_5(u) \left[\frac{1}{2\pi^2}\int_{x_i}^1 dx x\bar{x}(s-\tilde{m}_c^2)
-\left(\frac{2m_c\langle\bar{q}q\rangle}{3}-\frac{m_c^3\langle\bar{q}g_s \sigma Gq\rangle}{6T^4}\right)\delta(s-m_c^2)\right]\nonumber\\
&&\exp\left(-\frac{s+m_c^2+u\bar{u}m_\pi^2}{T^2} \right) -\frac{f_\pi m_\pi^2 m_c\langle\bar{q}g_s\sigma Gq\rangle}{36(m_u+m_d)T^2} \int_0^1 du \varphi_5(u) \exp\left(-\frac{2m_c^2+u\bar{u}m_\pi^2}{T^2} \right)  \nonumber\\
&&-\frac{f_{3\pi}m_\pi^2}{2\pi^2}\int_{m_c^2}^{s_0}ds \int_{x_i}^1 dx \bar{x} \left[\frac{3}{4}+s\,\delta(s-\tilde{m}_c^2) \right]\exp\left(-\frac{s+m_c^2}{T^2} \right)\nonumber\\
&&  -\frac{f_{3\pi}m_\pi^2}{2\pi^2 T^2}\int_{m_c^2}^{s_0}ds \int_{x_i}^1 dx x\bar{x} \,\tilde{m}_c^2\exp\left(-\frac{s+m_c^2}{T^2} \right) \nonumber\\
&&+\frac{f_{3\pi}m_\pi^2}{4\pi^2}\int_{m_c^2}^{s_0}ds \int_{x_i}^1 dx x \left[\frac{3}{2}-s\,\delta(s-\tilde{m}_c^2) \right]\exp\left(-\frac{s+m_c^2}{T^2} \right) \, ,
\end{eqnarray}
where $\lambda_{YD^*D^*}=\lambda_{Y}f_{D^*}^2M^2_{D^*}$, $\widetilde{M}_Y^2=\frac{M_Y^2}{4}$ and $x_i=\frac{m_c^2}{s}$. In numerical calculations, we take the  $C_{\pi}$  and $C_{Y}$  as free parameters, and search for the best values  to acquire  stable QCD sum rules.

\section{Numerical results and discussions}
We take  the standard values of the vacuum condensates,
$\langle
\bar{q}q \rangle=-(0.24\pm 0.01\, \rm{GeV})^3$,
$\langle\bar{q}g_s\sigma G q \rangle=m_0^2\langle \bar{q}q \rangle$,
$m_0^2=(0.8 \pm 0.1)\,\rm{GeV}^2$     at the   energy scale  $\mu=1\, \rm{GeV}$
\cite{SVZ79,Reinders85,Colangelo-Review},  and take the $\overline{MS}$  mass $m_{c}(m_c)=(1.275\pm0.025)\,\rm{GeV}$ from the Particle Data Group \cite{PDG}. We set $m_u=m_d=0$ and take account of
the energy-scale dependence of  the input parameters,
\begin{eqnarray}
\langle\bar{q}q \rangle(\mu)&=&\langle\bar{q}q \rangle({\rm 1GeV})\left[\frac{\alpha_{s}({\rm 1GeV})}{\alpha_{s}(\mu)}\right]^{\frac{12}{33-2n_f}}\, , \nonumber\\
 \langle\bar{q}g_s \sigma Gq \rangle(\mu)&=&\langle\bar{q}g_s \sigma Gq \rangle({\rm 1GeV})\left[\frac{\alpha_{s}({\rm 1GeV})}{\alpha_{s}(\mu)}\right]^{\frac{2}{33-2n_f}}\, , \nonumber\\
 m_c(\mu)&=&m_c(m_c)\left[\frac{\alpha_{s}(\mu)}{\alpha_{s}(m_c)}\right]^{\frac{12}{33-2n_f}} \, ,\nonumber\\
 \alpha_s(\mu)&=&\frac{1}{b_0t}\left[1-\frac{b_1}{b_0^2}\frac{\log t}{t} +\frac{b_1^2(\log^2{t}-\log{t}-1)+b_0b_2}{b_0^4t^2}\right]\, ,
\end{eqnarray}
  where   $t=\log \frac{\mu^2}{\Lambda_{QCD}^2}$, $b_0=\frac{33-2n_f}{12\pi}$, $b_1=\frac{153-19n_f}{24\pi^2}$, $b_2=\frac{2857-\frac{5033}{9}n_f+\frac{325}{27}n_f^2}{128\pi^3}$,  $\Lambda_{QCD}=210\,\rm{MeV}$, $292\,\rm{MeV}$  and  $332\,\rm{MeV}$ for the flavors  $n_f=5$, $4$ and $3$, respectively  \cite{PDG,Narison-mix}, and we choose  $n_f=4$.

 At the hadron side, we take the parameters  as $m_{\pi}=0.13957\,\rm{GeV}$,   $f_{\pi}=0.130\,\rm{GeV}$ \cite{Colangelo-Review},
 $M_{D^*}=2.01\,\rm{GeV}$, $f_{D^*}=263\,\rm{MeV}$, $s^0_{D^*}=6.4\,\rm{GeV}^2$  \cite{WangJHEP},
 $M_{Y}=4.48\,\rm{GeV}$,   $\lambda_{Y}=9.47 \times 10^{-2}\,\rm{GeV}^5$ \cite{WZG-Vector-NPB}, and  $f_{\pi}m^2_{\pi}/(m_u+m_d)=-2\langle \bar{q}q\rangle/f_{\pi}$ from the Gell-Mann-Oakes-Renner relation.

\begin{figure}
\centering
\includegraphics[totalheight=10cm,width=16cm]{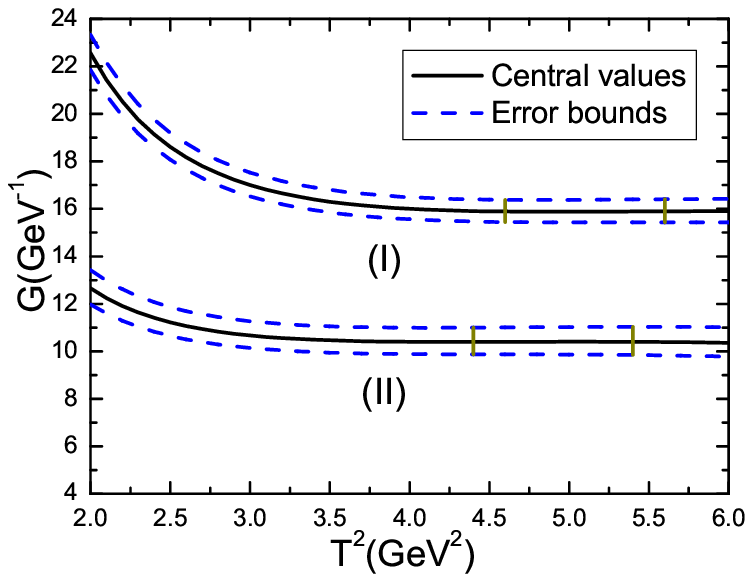}
  \caption{ The hadronic coupling constants with variations of the Borel parameters $T^2$, where the (I) and (II) denote the $G_{\pi}$ and $G_{Y}$, respectively, the regions between the two vertical lines are the Borel windows.  }\label{G-pi-X}
\end{figure}

In calculations, we fit the free parameters to be $C_{\pi}=0.00101(T^2-3.6\,\rm{GeV}^2)\,\rm{GeV}^4$ and
$C_{Y}=0.00089(T^2-3.2\,\rm{GeV}^2)\,\rm{GeV}^4$
  to acquire uniform flat Borel platforms  $T^2_{max}-T^2_{min}=1\,\rm{GeV}^2$ (just like in our previous works \cite{WZG-ZJX-Zc-Decay,WZG-Y4660-Decay,WZG-X4140-decay,
WZG-X4274-decay,WZG-Z4600-decay,WZG-Zcs3985-decay,WZG-Zcs4123-decay}), where the max and min represent the maximum and minimum values, respectively.
The Borel windows  are $T^2_{\pi}=(4.6-5.6)\,\rm{GeV}^2$ and $T^2_{Y}=(4.4-5.4)\,\rm{GeV}^2$, where  the subscripts $\pi$ and $Y$ represent the corresponding  channels,  the uncertainties  $\delta G_{\pi/Y}$ come from the Borel parameters $T^2$ are less than $0.01\, (\rm GeV^{-1})$. In Fig.\ref{G-pi-X}, we plot the hadronic coupling constants $G_{\pi}$ and $G_{Y}$ with variations of the Borel parameters. In the Borel windows, there appear very flat platforms indeed, it is reasonable and reliable to extract the $G_{\pi}$ and $G_{Y}$.

If we take  the symbol  $\xi$ to stand for the input parameters, then  the uncertainties   $\bar{\xi} \to \bar{\xi} +\delta \xi$ result in the uncertainties $\bar{\lambda}_{Y}\bar{f}_{D^*}\bar{f}_{\bar{D}^*}\bar{G}_{\pi/Y} \to \bar{\lambda}_{Y}\bar{f}_{D^*}\bar{f}_{\bar{D}^*}\bar{G}_{\pi/Y}
+\delta\,\bar{\lambda}_{Y}\bar{f}_{D^*}\bar{f}_{\bar{D}^*}\bar{G}_{\pi/Y}$, $\bar{C}_{\pi/Y} \to \bar{C}_{\pi/Y}+\delta C_{\pi/Y}$,
\begin{eqnarray}\label{Uncertainty-4}
\delta\,\bar{\lambda}_{Y}\bar{f}_{D^*}\bar{f}_{\bar{D}^*}\bar{G}_{\pi/Y} &=&\bar{\lambda}_{Y}\bar{f}_{D^*}\bar{f}_{\bar{D}^*}\bar{G}_{\pi/Y}\left( \frac{\delta f_{D^*}}{\bar{f}_{D^*}} +\frac{\delta f_{\bar{D}^*}}{\bar{f}_{\bar{D}^*}}+\frac{\delta \lambda_{Y}}{\bar{\lambda}_{Y}}+\frac{\delta G_{\pi/Y}}{\bar{G}_{\pi/Y}}\right)\, ,
\end{eqnarray}
where  the short overline \,$\bar{}$\, on all the input parameters   represents the central values.
In calculation, we observe that the uncertainties  $\delta C_{\pi/Y}$ are  very small, and set $\delta C_{\pi/Y}=0$ and $\frac{\delta f_{D^*}}{\bar{f}_{D^*}} =\frac{\delta f_{\bar{D}^*}}{\bar{f}_{\bar{D}^*}}=\frac{\delta \lambda_{Y}}{\bar{\lambda}_{Y}}=\frac{\delta G_{\pi/Y}}{\bar{G}_{\pi/Y}}
$ approximately. Now we obtain the hadronic coupling constants routinely,
\begin{eqnarray} \label{HCC-values}
G_{\pi} &=&15.9 \pm 0.5\,\rm{GeV}^{-1}\, , \nonumber\\
G_{Y} &=&10.4\pm 0.6\,\rm{GeV}^{-1}\, ,
\end{eqnarray}
by setting
\begin{eqnarray}\label{Uncertainty-5}
\delta\,\bar{\lambda}_{Y}\bar{f}_{D^*}\bar{f}_{\bar{D}^*}\bar{G}_{\pi/Y}  &=&\bar{\lambda}_{Y}\bar{f}_{D^*}\bar{f}_{\bar{D}^*}\bar{G}_{\pi/Y}\frac{4\delta G_{\pi/Y}}{\bar{G}_{\pi/Y}}\, .
\end{eqnarray}

Then it is direct  to obtain the partial decay width by taking the hadron masses
$M_{D^{*-}} = 2.01026\,\rm{GeV}$, $M_{D^{*0}}= 2.00685\,\rm{GeV}$ and
$m_{\pi} = 0.13957\,\rm{GeV}$ from the Particle Data Group \cite{PDG} and $M_{Y(4500)} = 4.4691\,\rm{GeV}$
from the BESIII collaboration \cite{X4500-BESIII},
\begin{eqnarray} \label{Partial-with}
\Gamma\left(Y(4500)\to D^*\bar{D}^* \pi^+\right)&=&\frac{1}{24\pi M_Y} \int dk^2 (2\pi)^4\delta^4(p^\prime-k-p)\frac{d^3 \vec{k}}{(2\pi)^3 2k_0}\frac{d^3 \vec{p}}{(2\pi)^3 2p_0} \nonumber\\
&& (2\pi)^4\delta^4(k-q-r)\frac{d^3 \vec{q}}{(2\pi)^3 2q_0}\frac{d^3 \vec{r}}{(2\pi)^3 2r_0}\Sigma |T|^2 \nonumber\\
&=&6.43^{+0.80}_{-0.76}\,\rm{MeV}\, ,
\end{eqnarray}
where $T=\langle Y_c(p^\prime)|\bar{D}^*(p)D^*(q)\pi(r)\rangle$ defined in Eq.\eqref{define-G-pi-X}.

The partial decay width  $\Gamma\left(Y(4500)\to D^*\bar{D}^* \pi^+\right) =6.43^{+0.80}_{-0.76}\,\rm{MeV}$ is much smaller than the total width  $\Gamma=246.3\pm 36.7\pm 9.4\,\rm{MeV}$ from the BESIII collaboration \cite{X4500-BESIII},  which is consistent with our naive expectation that the main decay channels of the vector tetraquark states are two-body strong decays $Y\to D\bar{D}$, $D^*\bar{D}^*$, $D\bar{D}^*$, $D^*\bar{D}$, $J/\psi \pi$, $\eta_c\rho$.  The observations of the $Y(4500)$ in the channels $ D\bar{D}$, $D^*\bar{D}^*$, $D\bar{D}^*$, $D^*\bar{D}$, $J/\psi \pi$, $\eta_c\rho$  would shed light on the nature of the $Y(4500)$, and we would explore those two-body strong decays in our next work in a comprehensive way.

We choose the process $Y(4500)\to D^{*-}D^{*0}\pi^+$ to explore whether or not the four-hadron coupling constants can be calculated directly using the (light-cone) QCD sum rules, as this  process is not expected to be the dominant decay channel, which only servers as a powerful constraint to examine the calculations, i.e. the partial decay width should be small enough to satisfy  the BESIII experimental data. We should admit that it would be better to find a tetraquark candidate, whose dominant decay mode is the three-body strong decay, to examine the present approach (or procedure), however, at the present time, we cannot find such a tetraquark candidate.
 In short,  the present work  supports assigning the $Y(4500)$ to be the $[uc]_{\tilde{A}}[\overline{uc}]_{V}+[uc]_{V}[\overline{uc}]_{\tilde{A}}+[dc]_{\tilde{A}}[\overline{dc}]_{V}
+[dc]_{V}[\overline{dc}]_{\tilde{A}}$ hidden-charm tetraquark state with the quantum numbers  $J^{PC}=1^{--}$.
It is the first time to use the light-cone QCD sum rules to study the four-hadron coupling constants, the approach can be used to explore the
$Y\to J/\psi \pi^+\pi^-$, $\psi^\prime \pi^+\pi^-$, $J/\psi K^+K^-$, $h_c \pi^+\pi^-$, $D^0D^{*-}\pi^+$, and diagnose  the nature of the $X$, $Y$ and $Z$ states.

\section{Conclusion}
 In this work, we tentatively assign the $Y(4500)$ as the $[uc]_{\tilde{A}}[\overline{uc}]_{V}+[uc]_{V}[\overline{uc}]_{\tilde{A}}+[dc]_{\tilde{A}}[\overline{dc}]_{V}
+[dc]_{V}[\overline{dc}]_{\tilde{A}}$ tetraquark state with the quantum numbers $J^{PC}=1^{--}$, and   extend our previous works to study the three-body strong decay $Y(4500)\to D^{*-}D^{*0}\pi^+$ with the light-cone QCD sum rules, the partial width is consistent with the experimental data from the BESIII collaboration. It is the first time to use the light-cone QCD sum rules to study the four-hadron coupling constants, we choose the process $Y(4500)\to D^{*-}D^{*0}\pi^+$ to explore whether or not the (light-cone) QCD sum rules can be used to calculate the four-hadron coupling constants directly, as the process is not the main decay channel, which servers as a powerful constraint to testify the approach, i.e. the partial decay width should be small enough to be match the experimental data.
The approach can be used to investigate the three-body strong decays $X/Y\to J/\psi \pi^+\pi^-$, $\psi^\prime \pi^+\pi^-$, $J/\psi K^+K^-$, $h_c \pi^+\pi^-$, $D^0D^{*-}\pi^+$ directly,  and shed light on the nature of the $X$, $Y$ and $Z$ states.

\section*{Acknowledgements}
This  work is supported by National Natural Science Foundation, Grant Number  12175068.

\end{document}